\documentclass{article}
\usepackage{cite}
\usepackage{amsmath,amssymb,amsfonts, blkarray,algorithmic}
\usepackage{algorithmic}
\usepackage{graphicx}
\usepackage{textcomp}
\usepackage{wrapfig}
\usepackage[flushleft]{threeparttable}
\usepackage{subcaption}
\def\BibTeX{{\rm B\kern-.05em{\sc i\kern-.025em b}\kern-.08em
    T\kern-.1667em\lower.7ex\hbox{E}\kern-.125emX}}
\begin{document}
\date{}

\title{Magnetic Sensor Compensation Utilizing Factor Graph Estimation}
\author{Frederic W. Lathrop, Clark N. Taylor, and Aaron Nielsen}
\thanks{F. W. Lathrop is with the Air Force Institute of Technology, Wright-Patterson Air Force Base, OH 45433 (email: frederic.lathrop@au.af.edu). }
\thanks{C. N. Taylor is with the Electrical and Computer Engineering Department, Air Force Institute of Technology, Wright-Patterson Air Force Base, OH 45433 (email: clark.taylor.3@au.af.edu). }
\thanks{A. P. Nielsen is with the Electrical and Computer Engineering Department, Air Force Institute of Technology, Wright-Patterson Air Force Base, OH 45433 (email: aaron.nielsen.2@au.af.edu).}

\maketitle

\begin{abstract}
Recently, there has been significant interest in the ability to navigate without GPS using the magnetic anomaly field of the Earth (magnav).  One of the key technical bottlenecks to achieving magnav is obtaining an accurate magnetic sensor calibration, taking into account own-ship and sensor effects. The Tolles-Lawson magnetic calibration method was developed when airborne magnetic survey aircraft were first employed 80 years ago.  In this paper, we present a magnetic calibration algorithm based on a factor graph optimization using inertial measurements as well as inputs from both a vector and scalar magnetometer.  The factor graph is well suited for combining multiple sensor inputs and allows accurate calibrations in the presence of large permanent moments as well as estimating a time-varying external magnetic field during the calibration, two problems that are difficult to solve with previous approaches. The ability to accurately calibrate a magnetic sensor in flight (i.e. with the presence of a large platform field and a varying Earth field due to movement) will allow greater flexibility in sensor mounting locations for magnetic anomaly navigation. 
\end{abstract}



\section{Introduction}
\label{sec:introduction}
Attempting to utilize magnetic anomaly signals for navigation has been researched for several decades. Early efforts did not provide an absolute positioning system, but explored the possibility of utilizing magnetic anomaly signals to aid velocity calculations for dead reckoning navigation in submarines \cite{VelocityAiding}.  In addition to being one of the first to use magnetic anomalies to aid navigation, this work also recognized the importance of platform calibration when using these measurements. Later efforts discussed aircraft positioning using magnetic fields and addressed sensor placement and calibration \cite{GeoMagNav}. Early experimental results of aircraft positioning using magnetic measurements to aid a dead reckoning navigation filter were presented in 2006 \cite{MagFieldVariations} and more recently absolute positioning of an aircraft utilizing real flight data has been demonstrated \cite{MagNav}. Research has also focused on improving calibration methods \cite{OnlineCal}. This paper addresses a critical need for enabling magnetic anomaly navigation: accurate platform calibration.  

 Tolles-Lawson is the foundation still used today for platform compensation for most aerial magnetic surveys \cite{TollesLaswon}. Survey aircraft normally mount magnetic sensors in booms or other devices to physically separate the sensors from the aircraft disturbance field.  Under these conditions, the Tolles-Lawson method works well to remove the remaining aircraft disturbances from the measurements, allowing accurate survey maps to be generated. The Tolles-Lawson compensation method uses an aircraft magnetic disturbance model that identifies the three primary types of disturbances: permanent moments,  induced moments, and eddy currents \cite{MAD}.  The method solves for 18 coefficients (3 permanent, 6 induced, 9 eddy current) that are used with the vector magnetometer providing the total magnetic field orientation in order to determine the aircraft scalar disturbance at each time step.  The major simplifying assumption made in order to solve for the coefficients is the disturbance field is much smaller than the earth field, and therefore orientation information derived from the vector magnetometer is not significantly affected by the disturbance and a first order Taylor series expansion is used to linearize the problem \cite{gnadt2022derivation}. Additionally, Tolles-Lawson is used to compensate the scalar measurement and does not provide any calibration information for the vector sensor used.

There have been numerous publications focusing on vector sensor calibration over the past two decades. The TWOSTEP method was one of the first algorithms that enabled attitude-independent calibration is still frequently used as a performance benchmark \cite{TWOSTEP}. Papers published over the two decades following TWOSTEP can be characterized in two primary groups: those using vector sensor measurements only \cite{TWOSTEP, FastCal, wu2015calibration} and methods incorporating inertial sensors in addition to vector sensors \cite{MAVBE, kok2016magnetometer, papafotis2020accelerometer}. 

Factor graphs present a flexible optimization method suitable for non-linear problems such as magnetic calibrations. The flexibility also allows the factor graph to be constructed such that it calibrates both vector and scalar sensors, unlike previously described methods that calibrate either a scalar or vector sensor, but not both. Originally introduced in 2001 \cite{FactorGraph}, factor graphs have been used in robotics, simultaneous localization and mapping (SLAM), and are equally suited for optimizing platform compensation.    Many contemporary calibration methods solve a maximum likelihood or utilize an extended Kalman filter (EKF), however factor graphs have been shown to have advantages when compared to EKFs for estimation problems\cite{wen2021factor}.    

Our contribution with this paper is to demonstrate a calibration routine utilizing a factor graph optimization incorporating inertial information.  While the current algorithm is limited to hard iron calibration, it has the benefit when compared to other calibration methods that it has improved performance in large hard iron disturbance fields and allows a time varying external magnetic field during the calibration. 

This paper will present an overview of our calibration method using a factor graph optimization.  Section \ref{section:Hard Iron Calibration} will discuss the magnetic models used as well as the structure of the factor graph.  Section \ref{section:performance} will evaluate the algorithm's performance when compared to Alonso and Shuster's TWOSTEP method as well as a Tolles-Lawson calibration. The comparison is first conducted using a Monte Carlo simulation followed by real data collected in a controlled experiment.  Section \ref{section:Conclusion} concludes the paper.  

\section{Calibration in Large Body Fields} \label{section:Hard Iron Calibration}

The notation used throughout the remainder of the paper is summarized below:

\subsection{Notation} \label{section:Notation}

\begin{enumerate}
    \item \textbf{Bold upper case (T)}: matrix
    \item \textbf{bold lower case (h)}: vector
    \item lower case (m): scalar
    \item Tilde ($\Tilde{\ \ }$): measured value
    \item Circumflex ($\hat{\ \ }$): estimated value 
     
    \item Coordinate Frames: The two coordinate frames referenced in this paper will be the navigation frame and the body frame.  The navigation frame is fixed throughout the simulation and utilizes a North-East-Down (NED) right hand coordinate system.  The body frame is a right hand coordinate system fixed to the platform such that the x-y-z coordinates correspond with roll-pitch-yaw of the platform with the y-axis pointing to the right and the z-axis pointing down. 
    \item $\boldsymbol{e}^n$ : Earth's total magnetic field in the navigation frame
    \item Superscripts: used to indicate the coordinate frame of a given vector.  When followed by an integer, the integer represents which time step the frame is in. The two primary frames referenced will be the navigation frame ($n$) and the body frame ($b$). \label{note:superscript}
    \item Direction Cosine Matrices (DCM): $\boldsymbol{C}_a^b$ represents a right multiply (column space) 3x3 DCM that transforms a vector from the $a$ frame to the $b$. This notation combined with item \ref{note:superscript} above, means $\boldsymbol{C}_n^{bk}$ represents the rotation from the navigation frame to the body frame at time step k. The Earth's magnetic field in the body frame at time step 0 is represented as $\boldsymbol{e}^{b0} = \boldsymbol{C}_n^{b0}\boldsymbol{e}^n$

\end{enumerate}

\subsection{Magnetic Model}\label{section:meas model}

A simplified mathematical model of the vector magnetic measurements was used following the error sources frequently used as summarized by \cite{MagCal_survey}.  The primary sources of error in this model are:
\begin{itemize}
    \item Sensor bias $(\boldsymbol{h}_{vec})$: a constant bias on each vector component 
    \item Hard iron distortion $(\boldsymbol{h}_{hi})$: changes in the sensor magnetic field due to magnetic materials attached to the platform or body frame.
    \item Soft iron distortion $(\boldsymbol{T}_{si})$: Distortion in the sensed magnetic field due to material properties of the platform, modeled as a 3x3 positive definite symmetric matrix.  The amount of distortion depends on the orientation and strength of the external field. 
    \item Non-orthogonality $(\boldsymbol{T}_{ortho})$: While generally small, imperfections in mechanical characteristics of sensors as well as other contributing factors, such as imperfect installations, resulting in the component values of the sensor not being truly orthogonal to each other or other sensors. Modeled as a 3x3 matrix with 3 degrees of freedom, as shown in (\ref{eq:T_ortho}).  
    \item Scale Factor $(\boldsymbol{T}_{scale})$: gain error in each axis of the magnetometer; 3x3 diagonal matrix. 
    \item Noise $(\boldsymbol{\epsilon})$: Random errors modeled as zero mean, white noise $\left(\mathcal{N}(0,\sigma^2)\right)$.
\end{itemize}

Combining the different error components leads to the following model for the magnetic measurements for a given external field, $\boldsymbol{e}^n$: 

\begin{align}
    \boldsymbol{m}_{vec} =& \begin{bmatrix}
        m_x \\ m_y \\ m_z \end{bmatrix}  \\ =& \boldsymbol{T}_{scale}\boldsymbol{T}_{ortho}\left(\boldsymbol{T}_{si}\boldsymbol{C}_n^b\boldsymbol{e}^n + \boldsymbol{h}_{hi}\right) + \boldsymbol{h}_{vec} + \boldsymbol{\epsilon}_{vec} 
    \label{eq:meas_model_orig} 
\end{align}

\begin{figure}[h]
    \centering
    \includegraphics{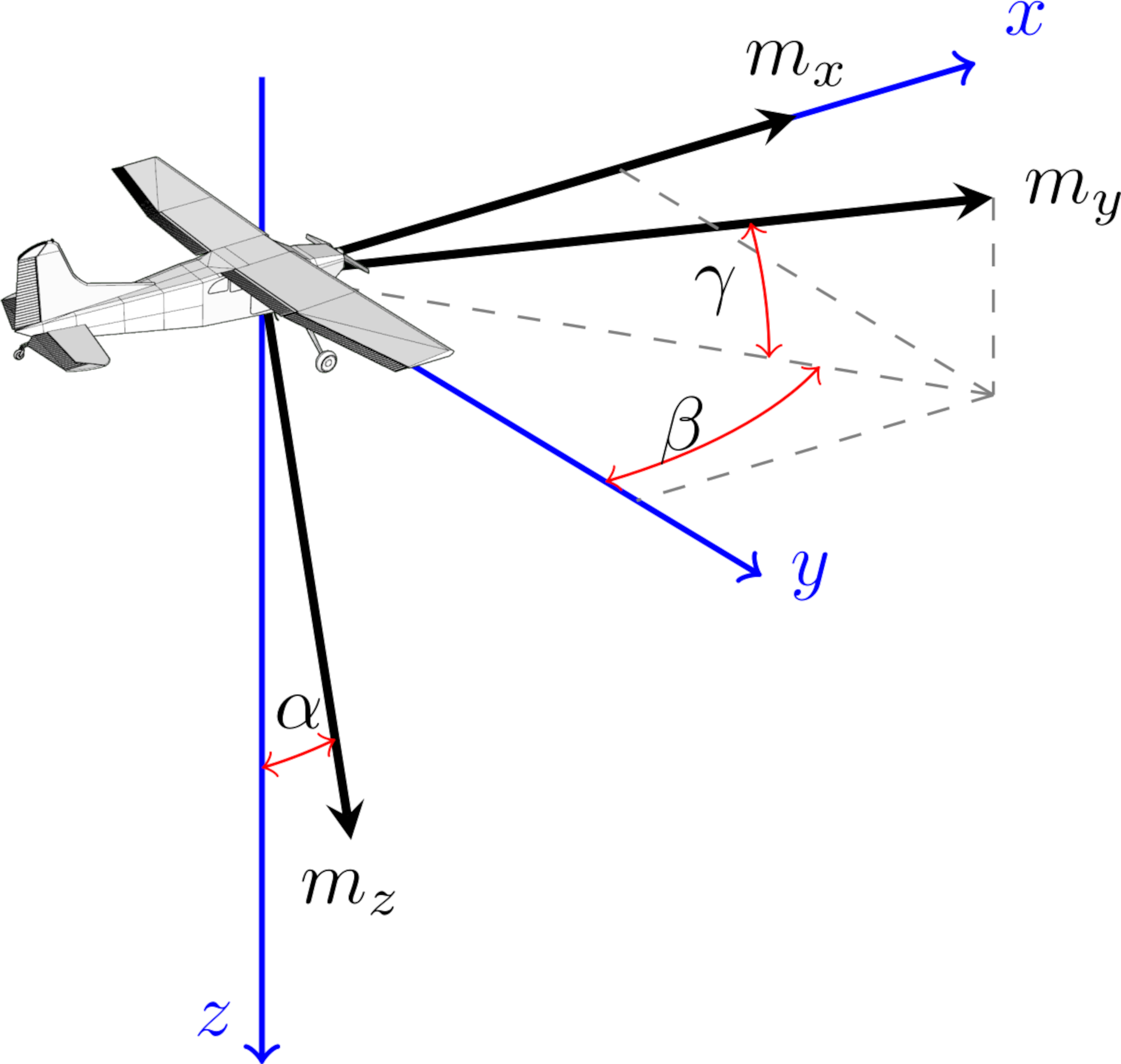}
    \caption{Non-Orthogonal Measurements in Sensor Frame}
    \label{fig:non_ortho}
\end{figure}

The scale factor, $\boldsymbol{T}_{scale}$ is a diagonal 3x3 matrix with each value corresponding to the respective axis scale value and the non-orthogonal errors are modeled following the method provided by \cite{FastCal}, defined by three non-orthogonal angles while assuming the x-axis of the sensor is aligned with the truth, as depicted in Fig. \ref{fig:non_ortho} and (\ref{eq:T_ortho}).

\begin{align}
    \boldsymbol{T}_{scale} &= \begin{bmatrix} k_x & 0 & 0 \\ 0 & k_y & 0 \\ 0 & 0 & k_z \end{bmatrix}
    \label{eq:T_scale} \\
    \boldsymbol{T}_{orth} &= \begin{bmatrix} 1 & 0 & 0 \\ \sin(\beta)\cos(\gamma) & \cos(\beta)\cos(\gamma) &  \sin(\gamma)\\ \sin(\alpha) & 0 & \cos(\alpha)
    \label{eq:T_ortho}
    \end{bmatrix}
\end{align}

Soft iron effects, $\boldsymbol{T}_{si}$, are commonly modeled as a positive definite symmetric matrix \cite{OnlineCal} and the same model is used in this paper. Soft iron effects were included in the magnetic measurement models to improve realism, however the factor graph only estimates $\boldsymbol{h}_{hi}$, the hard iron or permanent moments. The positive definite symmetric soft iron distortion matrix is a 3x3 element matrix with six unique terms in the following form: 

\begin{equation}
    \boldsymbol{T}_{si} = \begin{bmatrix}
        a & b & c \\
        b & d & e \\
        c & e & f \end{bmatrix}
        \label{eq:Soft Iron}
\end{equation}

Simplifying (\ref{eq:meas_model_orig}) and letting $\boldsymbol{T} = \boldsymbol{T}_{scale}\boldsymbol{T}_{ortho}$ results in

\begin{equation}
    \boldsymbol{m}_{vec} = \boldsymbol{T}\left(\boldsymbol{T}_{si}\boldsymbol{C}_n^b\boldsymbol{e}^n + \boldsymbol{h}_{hi}\right) + \boldsymbol{h}_{vec} + \boldsymbol{\epsilon}_{vec}
    \label{eq:meas_vec}
\end{equation}

where

\begin{equation}
        \boldsymbol{T} = \begin{bmatrix} k_x & 0 & 0 \\ \sin(\beta)\cos(\gamma) & k_y\cos(\beta)\cos(\gamma) &  \sin(\gamma)\\ \sin(\alpha) & 0 & k_z\cos(\alpha) \end{bmatrix}
        \label{eq:T}
\end{equation}

The scalar sensor will have the same errors as the vector sensor due to hard iron ($\boldsymbol{h}_{hi}$) and soft iron effects ($\boldsymbol{T}_{si}$), however $\boldsymbol{T}$ does not apply to scalar measurements. The model used for the scalar measurements is shown in (\ref{eq:meas_scalar}). This simplified model does not account for all possible errors, most notably heading errors associated with optically pumped magnetometers.  These errors can be minimized with optimized sensor orientations and were not modeled in simulation.

\begin{align}
    m_{scalar} &= \begin{Vmatrix} \boldsymbol{T}_{si}\boldsymbol{C}_n^b\boldsymbol{e}^n + \boldsymbol{h}_{hi} \end{Vmatrix} +  \epsilon_{scalar}
    \label{eq:meas_scalar}
\end{align}

\subsection{Factor Graph}
\begin{figure*}[h]
    \centering
    \includegraphics[width=.7\textwidth]{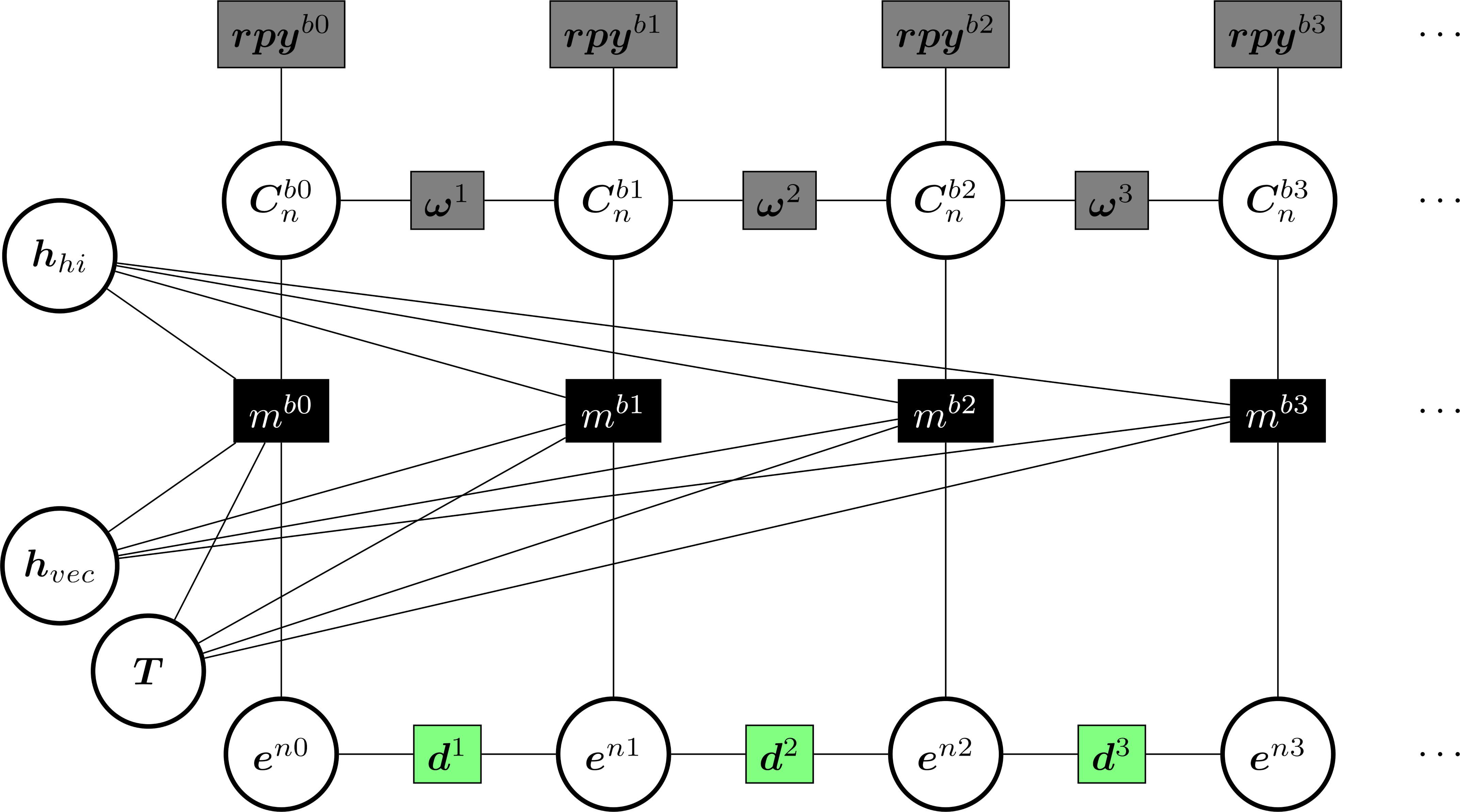}
    \caption{Time Variant Magnetic Calibration Factor Graph}
    \label{fig:FG}
\end{figure*}

A factor graph is a way to build a graphical model to visualize system interactions and set up a non-linear least squares optimization. A factor graph is a ``a bipartite graphical representation of a mathematical relation" \cite{FactorGraph}.  This graphical representation consists of variable nodes and factor nodes.  Each unknown variable to be estimated is represented by a variable node, an open circle in the factor graph.  The local functions relating the variables are factor nodes and represented by squares nodes in the graphical form.  Lines connecting variable and function nodes are used to show relationships and indicate where corresponding terms in the equivalent matrix notation are necessary.  The factor graph is bipartite because variable nodes are only connected to factor nodes, and vice versa.  The factor graph for the first four time steps of our estimation problem is shown in Fig. \ref{fig:FG}. 

The variable nodes in Fig. \ref{fig:FG} depict the unknowns in the system: hard iron ($\boldsymbol{h}_{hi}$), vector bias ($\boldsymbol{h}_{vec}$), external magnetic field ($\boldsymbol{e}^{nk}$), scale factor and non-orthogonality errors ($\boldsymbol{T}$), and the orientation at each time step ($\boldsymbol{C}_{n}^{bi}$ nodes). Together, all the nodes create the state of the system, $\boldsymbol{x}$. For readability, the variable nodes in the state form are shown in their matrix form, however for implementation any matrix variable node is vectorized.  Additionally,  DCMs are converted to 3-element Rodrigues' vectors ($\mathcal{S}\mathcal{O}(3) \Rightarrow \mathcal{R}^3$), but shown in DCM notation for clarity. $\boldsymbol{T}$ in the state is the same as (\ref{eq:T}), and $\boldsymbol{T}^v$ is the vectorized form.  

\begin{align}
    \boldsymbol{x} &= \begin{bmatrix} 
        \boldsymbol{h}_{hi} \\
        \boldsymbol{h}_{vec} \\
        \boldsymbol{T}^v \\
        \boldsymbol{e}^{n0} \\
        \boldsymbol{e}^{n1} \\
        \vdots \\
        \boldsymbol{e}^{nk-1} \\
        \boldsymbol{e}^{nk} \\
        \boldsymbol{C}_{n}^{b0} \\
        \vdots \\
        \boldsymbol{C}_{n}^{bk} \end{bmatrix}
    \label{eq:x} \\
    \boldsymbol{T}^v &= \begin{bmatrix} k_x & k_y & k_z & \alpha & \beta & \gamma \end{bmatrix}^\top
    \label {eq:T_vec}
\end{align}

The factor nodes represent probabilistic constraints on the variable nodes and in our implementation are the weighted measurements from the magnetic sensors and the inertial measurements, shown in (\ref{eq:mag_meas})-(\ref{eq:rpy_meas}). The component values of the vector magnetometer sensor are $m_x$, $m_y$, and $m_z$ and $m_{scalar}$ is the scalar magnetometer measurement. The $\boldsymbol{\omega}^n$ factor nodes are comprised of the incremental attitude changes in the platform frame about the axis indicated: $\omega_x$, $\omega_y$, and $\omega_z$. The $\boldsymbol{rpy}^{bn}$ factors are the roll, pitch, and heading Euler angles of the platform: $\phi$, $\theta$, and $\psi$ respectively. The Euler angle measurements are assumed to be available from an inertial navigation system (INS) or attitude heading reference system (AHRS) and are transformed into a corresponding DCM for all calculations. The changes in the external magnetic field are captured by the $\boldsymbol{d}^n$ factor nodes. The covariance of the $\boldsymbol{d}^n$ factor nodes can be adjusted based on the confidence of having a steady state vs. changing external field. 

\begin{align}
    \boldsymbol{\tilde{m}}^{bi} &= \begin{bmatrix} \tilde{m}_x & \tilde{m}_y & \tilde{m}_z & \tilde{m}_{scalar} \end{bmatrix}^\top 
    \label{eq:mag_meas} \\
    \boldsymbol{\tilde{\omega}}^{bi} &= \begin{bmatrix} \tilde{\omega}_x & \tilde{\omega}_y & \tilde{\omega}_z   \end{bmatrix}^\top
    \label{eq:gyro_meas} \\
    \boldsymbol{\widetilde{rpy}}^{bi} &= \begin{bmatrix} \tilde{\phi} & \tilde{\theta} & \tilde{\psi}    
    \end{bmatrix}^\top \Rightarrow \boldsymbol{\tilde{C}}_n^{bi}
    \label{eq:rpy_meas} 
\end{align}

The residual vector, $\boldsymbol{y}$, is the difference between the sensor measurements and the estimated measurements based on the system state. This results in a column vector of residuals according to (\ref{eq:y}). Similar to the state vector in (\ref{eq:x}), rotations are depicted in their DCM format in (\ref{eq:y}) however the calculations are performed using rotation vectors. The elements in the residual vector are weighted based on the uncertainty of the corresponding measurements. The weighting of the change in external field can be tuned, dependent on the confidence of a static external field assumption, or the accuracy of the magnetic map when performing a map based calibration.   

\newlength{\ts} 
\setlength{\ts}{-.5pt}

The factor graph is used to create an equivalent matrix, $\boldsymbol{L}$, which is the Jacobian of the nonlinear least squares problem with one column for each variable node and one row for each factor node.  The $\boldsymbol{L}$ matrix is a sparse matrix with non-zero values representing the connections between factor and variable nodes from the graphical representation. 

\begin{align}
    \boldsymbol{y} &= \left[ \begin{array}{c}
           \boldsymbol{d}^1\\\vdots\\\boldsymbol{d}^k \\
        \tilde{\boldsymbol{\omega}}^1-\boldsymbol{\omega}^1\\\vdots\\\tilde{\boldsymbol{\omega}}^k-\boldsymbol{\omega}^k\\
        \tilde{\boldsymbol{rpy}}^{b0} - \boldsymbol{rpy}^{b0}\\\vdots\\\tilde{\boldsymbol{rpy}}^{bk} - \boldsymbol{rpy}^{bk}\\[\ts]
        \\\tilde{\boldsymbol{m}}^0 - \boldsymbol{m}^0\\[\ts] \\\vdots\\[\ts] \\\tilde{\boldsymbol{m}}^ - \boldsymbol{m}^k\\[\ts]
    \end{array}\right] \\ 
    &= 
    \left[ \begin{array}{c}
         \hat{\boldsymbol{e}}^{n1} - \hat{\boldsymbol{e}}^{n0} \\
         \vdots \\
         \hat{\boldsymbol{e}}^{nk} - \hat{\boldsymbol{e}}^{nk-1} \\
         \boldsymbol{\tilde{\omega}}^1 - \hat{\boldsymbol{C}}_{n}^{b1} \hat{\boldsymbol{C}}_{b0}^{n}\\
         \vdots \\
         \boldsymbol{\tilde{\omega}}^k - \hat{\boldsymbol{C}}_{n}^{bk} \hat{\boldsymbol{C}}_{bk-1}^{n} \\
         \boldsymbol{\tilde{C}}_n^{b0} - \hat{\boldsymbol{C}}_n^{b0} \\
         \vdots \\
         \boldsymbol{\tilde{C}}_n^{bn} - \hat{\boldsymbol{C}}_n^{bn} \\
         \boldsymbol{\tilde{m}}^0 - \begin{bmatrix}\hat{\boldsymbol{h}}_{hi} + \hat{\boldsymbol{T}}\hat{\boldsymbol{C}}_n^{b0}\hat{\boldsymbol{e}}^{n0} + \hat{\boldsymbol{h}}_{vec}\\ \|\hat{\boldsymbol{h}}_{hi} + \hat{\boldsymbol{T}}\hat{\boldsymbol{C}}_n^{b0}\hat{\boldsymbol{e}}^{n0}\| \end{bmatrix} \\
         \vdots \\
         \boldsymbol{\tilde{m}}^k - \begin{bmatrix} \hat{\boldsymbol{h}}_{hi} + \hat{\boldsymbol{T}}\hat{\boldsymbol{C}}_{n}^{bk}\hat{\boldsymbol{e}}^{nk} + \hat{\boldsymbol{h}}_vec\\ \|\hat{\boldsymbol{h}}_{hi} + \hat{\boldsymbol{T}}\hat{\boldsymbol{C}}_{n}^{bk}\hat{\boldsymbol{e}}^{nk}\| \end{bmatrix} \\
    \end{array} \right] 
    \label{eq:y}
\end{align}

Measurement weighting is accomplished by a form of \it whitening \rm by pre-multiplying the sub-matrices $\boldsymbol{L}$ and the residuals in (\ref{eq:y}) by $\sum_i^{-1/2}$, where $\sum_i$ is the measurement covariance.  The scalar magnetometer measurements, for example, are divided by the measurement standard deviation.  \it Whitening \rm has the effect of making all values in the residual vector, $\boldsymbol{y}$, and the Jacobian matrix, $\boldsymbol{L}$, unit-less and allows measurements of different units (e.g. nanoTesla and radians) to be combined into a single cost function \cite{dellaert2017factor}. This pre-multiplication is not shown in (\ref{eq:y}) to improve readability.     

The $\boldsymbol{L}$, $\boldsymbol{x}$, and $\boldsymbol{y}$ matrices formed from the factor graph are then used in a Gauss-Newton optimization routine to determine the calibration parameters as described in the following sub-section.

\subsection{Batch Optimization Method}
\label{sec:optimization}
An iterative Gauss-Newton batch optimization algorithm is used to solve for the system state values.  Given an initial estimate for the state values, $\boldsymbol{x}_0$, an update is found by solving 
 the least-squares problem shown in (\ref{eq:gauss-newton}).  

\begin{equation}
    \Delta\boldsymbol{x} = \underset{\Delta}{\arg\min}\|\boldsymbol{L}\Delta-\boldsymbol{y}\|^2\Bigr\rvert_{\boldsymbol{x}=\boldsymbol{x}_0}
    \label{eq:gauss-newton}
\end{equation}

The least-squares problem is solved using the Moore-Penrose pseudo-inverse approach, which results in $\Delta x$:   

\begin{align}
    \boldsymbol{L}\Delta\boldsymbol{x} &\approx y \\
    \boldsymbol{L}^\top \boldsymbol{L}\Delta\boldsymbol{x} &\approx \boldsymbol{L}^\top\boldsymbol{y}  \\
    \Delta\boldsymbol{x} &\approx \left(\boldsymbol{L}^\top \boldsymbol{L}\right)^{-1}\boldsymbol{L}^\top\boldsymbol{y} 
    \label{eq:psuedo inverse}
\end{align}

The update is added to the initial estimate, $\boldsymbol{x}_0 + \Delta\boldsymbol{x} \to \boldsymbol{x}_1$ and the process is repeated until the solution converges, i.e. $\Delta\boldsymbol{x} \approx 0$.

\section{Performance Evaluation} \label{section:performance}

Our proposed methodology for estimating the magnetic measurement error sources was tested with both simulated and real data sets.  Simulation has the benefit of producing a true value with which the performance of the algorithm can be compared and allowing us to test the algorithm in different conditions.  Real data sets are used to prove the ultimate utility of our method.

\subsection{Simulation Setup} \label{sec:sim_setup}

\begin{figure}[htbp]
    \centering
    \begin{subfigure}[b]{.5\textwidth}
        \includegraphics{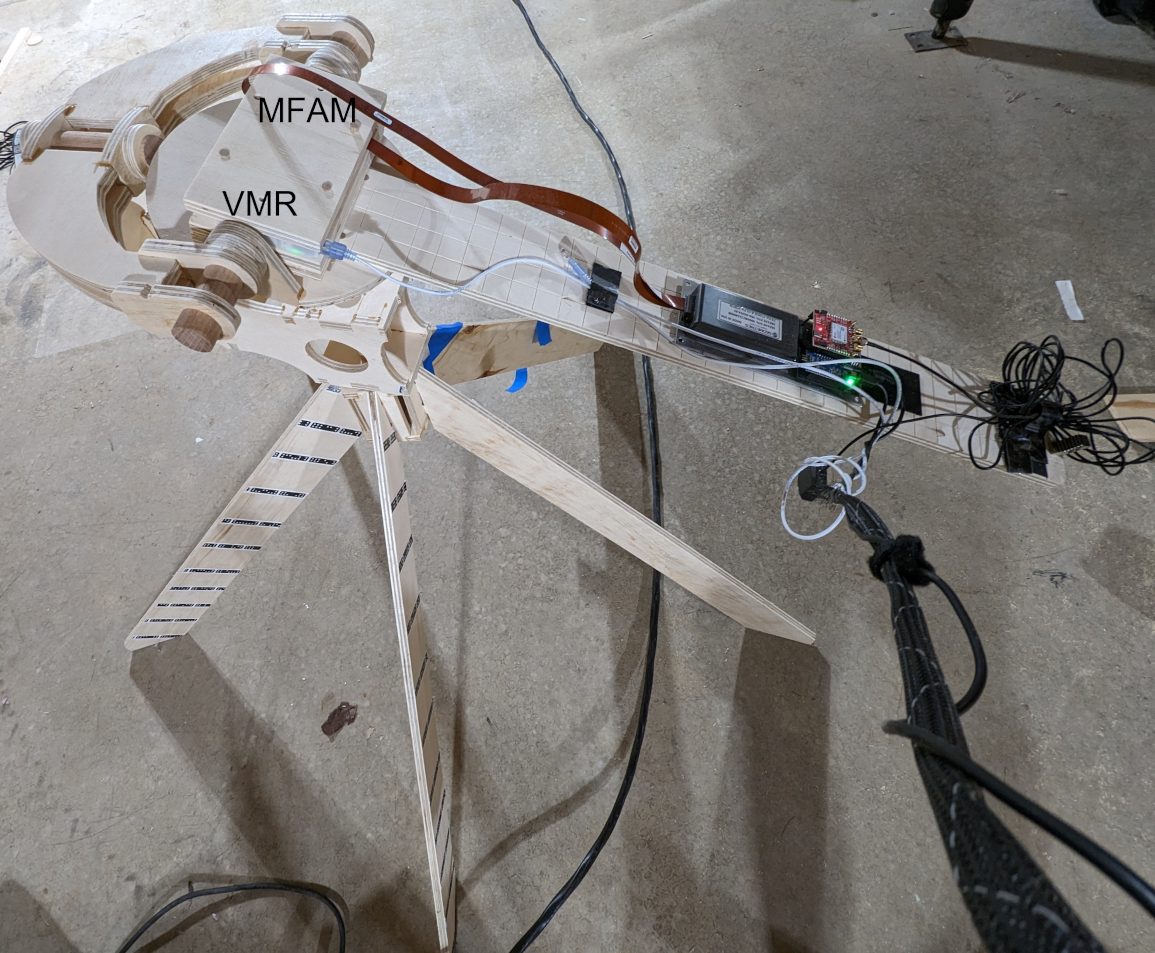}
        \caption{Test Platform}
        \label{fig:Tripod}
    \end{subfigure}
    \vfill
    \begin{subfigure}[b]{.5\textwidth}
        \includegraphics{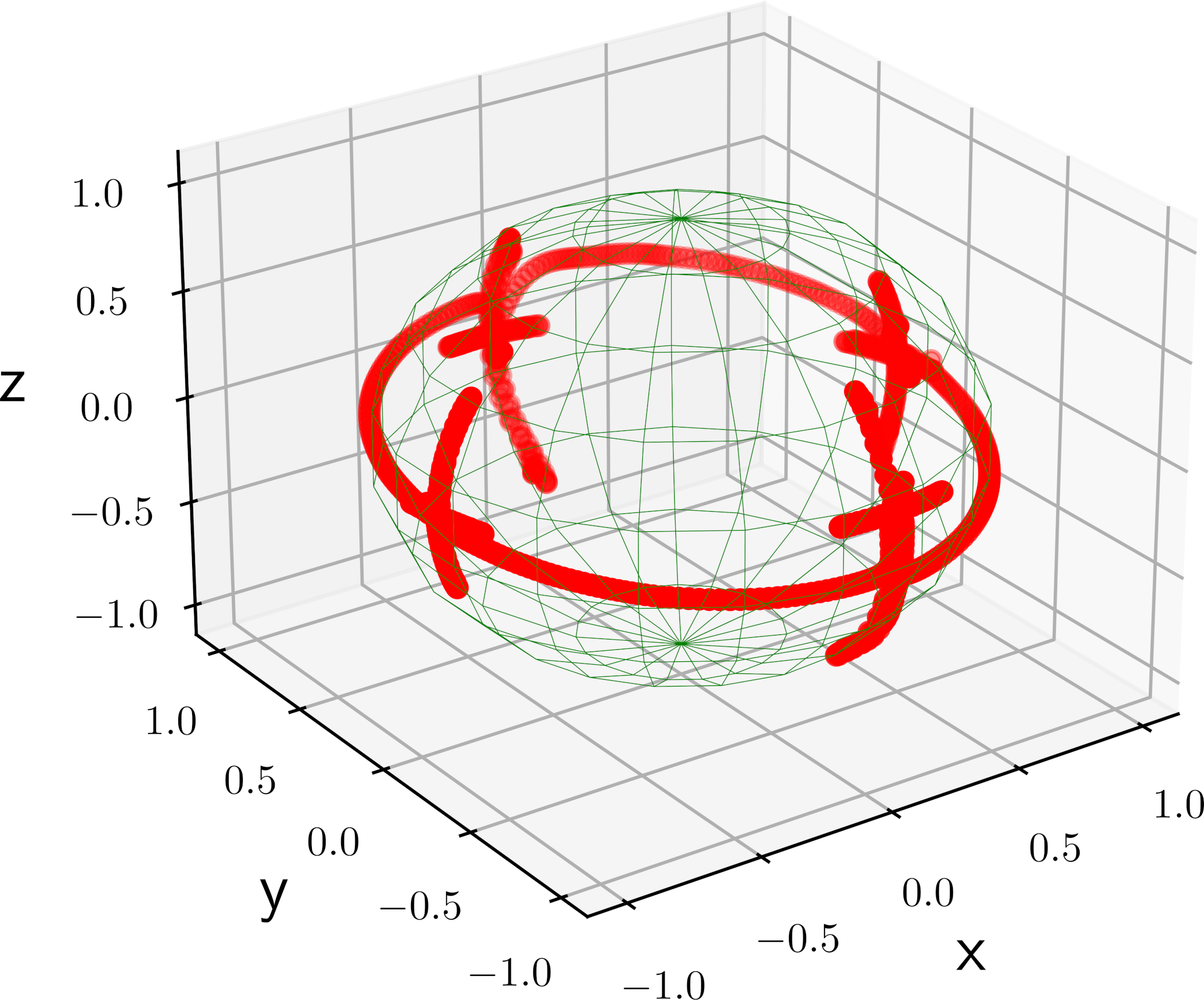}
        \caption{Calibration Profile}
        \label{fig:cal_profile}
    \end{subfigure}    
    \caption{}
    \label{fig:Test_Setup}
\end{figure}

Our simulation utilized a hybrid of real and simulated data.  Vector and scalar sensors were mounted on a tripod enabling the collection of magnetic and inertial data while changing the orientation of the platform, shown in Fig. \ref{fig:Tripod}.  The calibration profile we used consisted of a series of pitch, roll, and yaw doublets performed on four different headings roughly 90 degrees apart.  In Fig. \ref{fig:cal_profile}, we show the orientation of the tripod over time, demonstrating the pitch and yaw changes.  Roll oscillations were performed on each heading but are not depicted in Fig. \ref{fig:cal_profile}. The pitch, roll, and yaw doublets were limited in magnitude to be representative of maneuvers suitable for large aircraft.  The roll, pitch, and headings used for the simulation profiles were calculated from inertial measurements on the tripod during real data collections. 

After creating the attitude profile, an external magnetic field was simulated with a random orientation relative to the platform with an initial magnitude of 50,000 nT to represent Earth's core field.  Random walk noise was added to the simulated core field to demonstrate the algorithm's ability to handle a time-varying external field. IMU measurements were modeled with a non-zero bias, bias in run stability, and angular random walk (ARW) noise. Our previous work showed gyro quality did not have a significant impact on the results, therefore gyro noise parameters were used modeling a consumer grade MEMS gyro \cite{Mine}. Table \ref{tab:sim_param} summarizes the simulation parameters used for generating magnetic measurements according to \eqref{eq:meas_vec} and \eqref{eq:meas_scalar}. While the algorithm does not currently estimate soft iron errors, soft iron contributions were included in the simulation model.

\begin{table}[h]
   \renewcommand{\arraystretch}{1.3}
   \centering
   \caption{Simulation Parameters}
    \begin{threeparttable}
   \centering
   \begin{tabular}{|p{40pt}|p{65pt}|p{110pt}|}
   \hline
   \bf{Parameter} & \bf{Value} & \bf{Comments}  \\
   \hline\hline
   $\| \boldsymbol{e}^n_0 \|$  & $50000$ nT\tnote{1} & Initial external field strength.  \\
   \hline
   $\| \boldsymbol{h}_{hi} \|$  & $0 - 5000$ nT\tnote{1} & Platform hard iron contributions.  \\
   \hline
   $\| \boldsymbol{h}_{vec} \|$  & $1000$ nT\tnote{1} & constant bias on vector magnetometer  \\
   \hline
   $k_x$, $k_y$, $k_z$ &$\mathcal{N}(1,0.1)$ & \\
   \hline
   $\alpha$, $\beta$, $\gamma$ &$\mathcal{N}(0,0.01)$ rad&  \\
   \hline
   $\boldsymbol{T}_{si}$ & $\mathcal{I}_3 + \mathcal{N}(0,1e-5)$ & \\
   \hline
   \end{tabular}
   \begin{tablenotes}
    \small
    \item[1] Vector orientation randomly generated
   \end{tablenotes}
   
   \end{threeparttable}
   \label{tab:sim_param}
\end{table}

After generating the simulation truth data, measurements were created using the noise parameters listed in Table \ref{tab:meas_param}

\begin{table}[h]
   \renewcommand{\arraystretch}{1.3}
   \centering
   \caption{Measurement Parameters}
    \begin{threeparttable}
   \centering
   \begin{tabular}{|p{65pt}|p{65pt}|p{85pt}|}
   \hline
   \bf{Parameter} & \bf{Value} & \bf{Comments}  \\
   \hline\hline
   $\boldsymbol{\epsilon}_{vec}$  & $\mathcal{N}(0,5)$ nT &    \\
   \hline
   $\boldsymbol{\epsilon}_{scalar}$  & $\mathcal{N}(0,1)$ nT &   \\
   \hline
   $\boldsymbol{\epsilon}_{\omega}$ : $\begin{matrix}
     \text{bias instability} \\ \text{Random Walk}\end{matrix}$  & $\begin{matrix} \text{3.87e-5 $\text{rad}/\sqrt{\text{sec}}$} \\  \text{9.89e-5 rad/sec} \end{matrix}$ &Analog Devices ADIS16470 IMU \cite{website:ADIS} \\
   \hline
   $\boldsymbol{\epsilon}_{\phi, \theta, \psi}$ &$\mathcal{N}(0,0.043) \text{ rad}$ & AC20-181 \cite{AC20_181} \\
   \hline
   \end{tabular}
   \end{threeparttable}
   \label{tab:meas_param}
\end{table}

\subsection{Simulation Results}

100 Monte Carlo simulations were run to evaluate the factor graph's estimation accuracy.  Additional Monte Carlo simulations were conducted to compare hard iron estimate accuracy for varying magnitudes of hard iron between our factor graph method, the TWOSTEP method as implemented by \cite{TWOSTEPimp}, and a Tolles-Lawson calibration as described by \cite{gnadt2022derivation}. The four metrics used to assess estimation accuracy of the Factor Graph calibration are shown in \eqref{eq:RMSE}-\eqref{eq:epsilon_ortho}.  The error in hard iron estimation, $\epsilon_{hi}$, shown \eqref{eq:epsilon_h_hi} was used when comparing the three calibration algorithms. 

\begin{figure*}
    \centering
    \includegraphics[width=\textwidth]{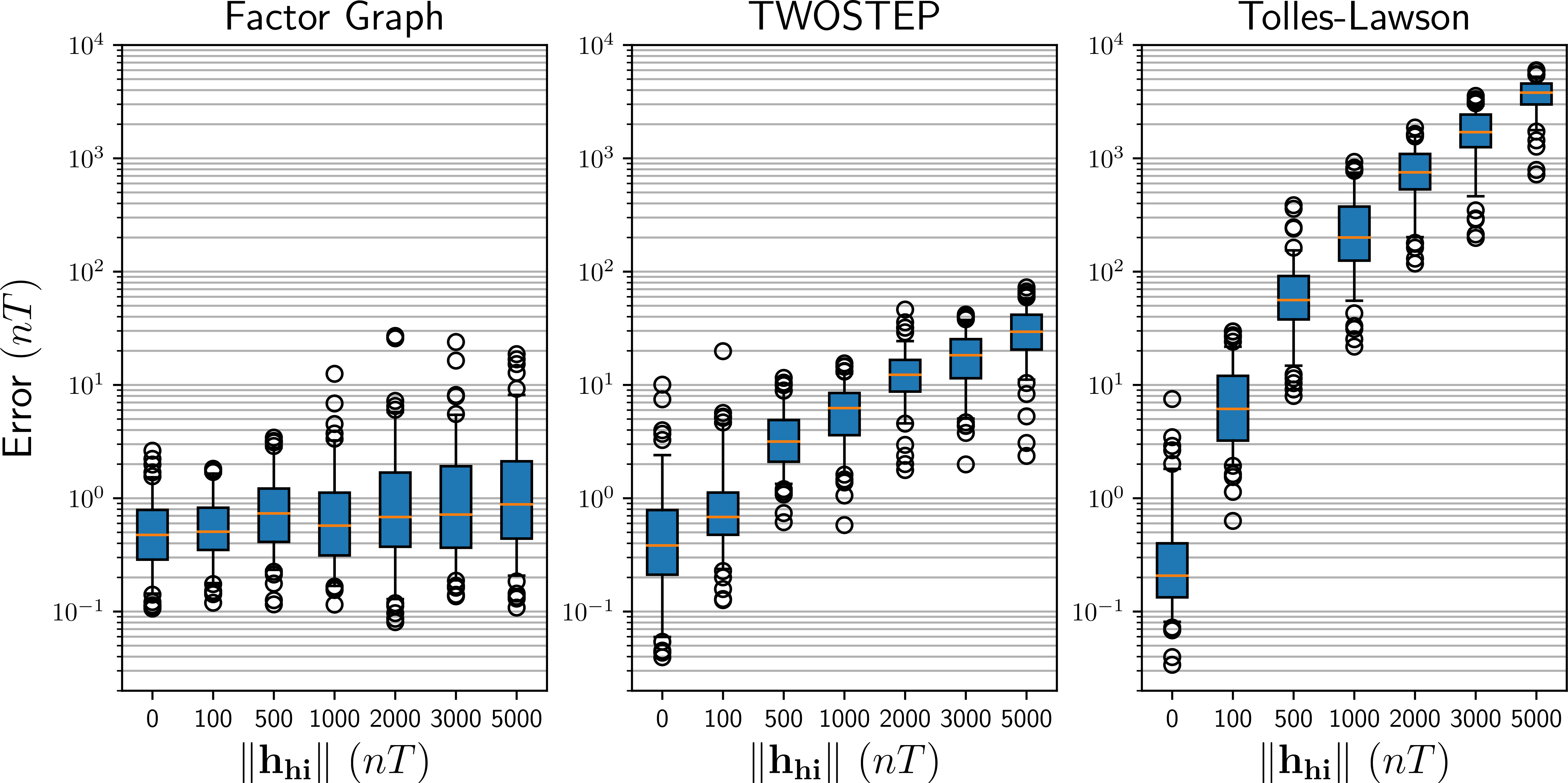}
    \caption{Hard Iron Magnitude Effect on Estimation Accuracy}
    \label{fig:h_hi_variation}
\end{figure*}

\begin{align}
    \label{eq:RMSE}
    RMSE &\equiv \sqrt{\frac{\Sigma_{i=1}^k\left(\|\boldsymbol{e}^n_i\| - \|\hat{\boldsymbol{e}}^n_i\|\right)^2}{k}}\\ 
    \label{eq:epsilon_h_hi}
    \epsilon_{hi} &\equiv \sqrt{\|\boldsymbol{h}_{hi} - \hat{\boldsymbol{h}}_{hi}\|^2} \\ 
    \label{eq:epsilon_scale}
    \epsilon_{scale} &\equiv \sqrt{\frac{(k_x - \hat{k}_x)^2 + (k_y - \hat{k}_y)^2 +(k_z - \hat{k}_z)^2}{3}} \\ 
    \label{eq:epsilon_ortho}
    \epsilon_{ortho} &\equiv \sqrt{\frac{(\alpha - \hat{\alpha})^2 + (\beta - \hat{\beta})^2 +(\gamma - \hat{\gamma})^2}{3}}  
\end{align}

\begin{figure}
    \centering
    \includegraphics{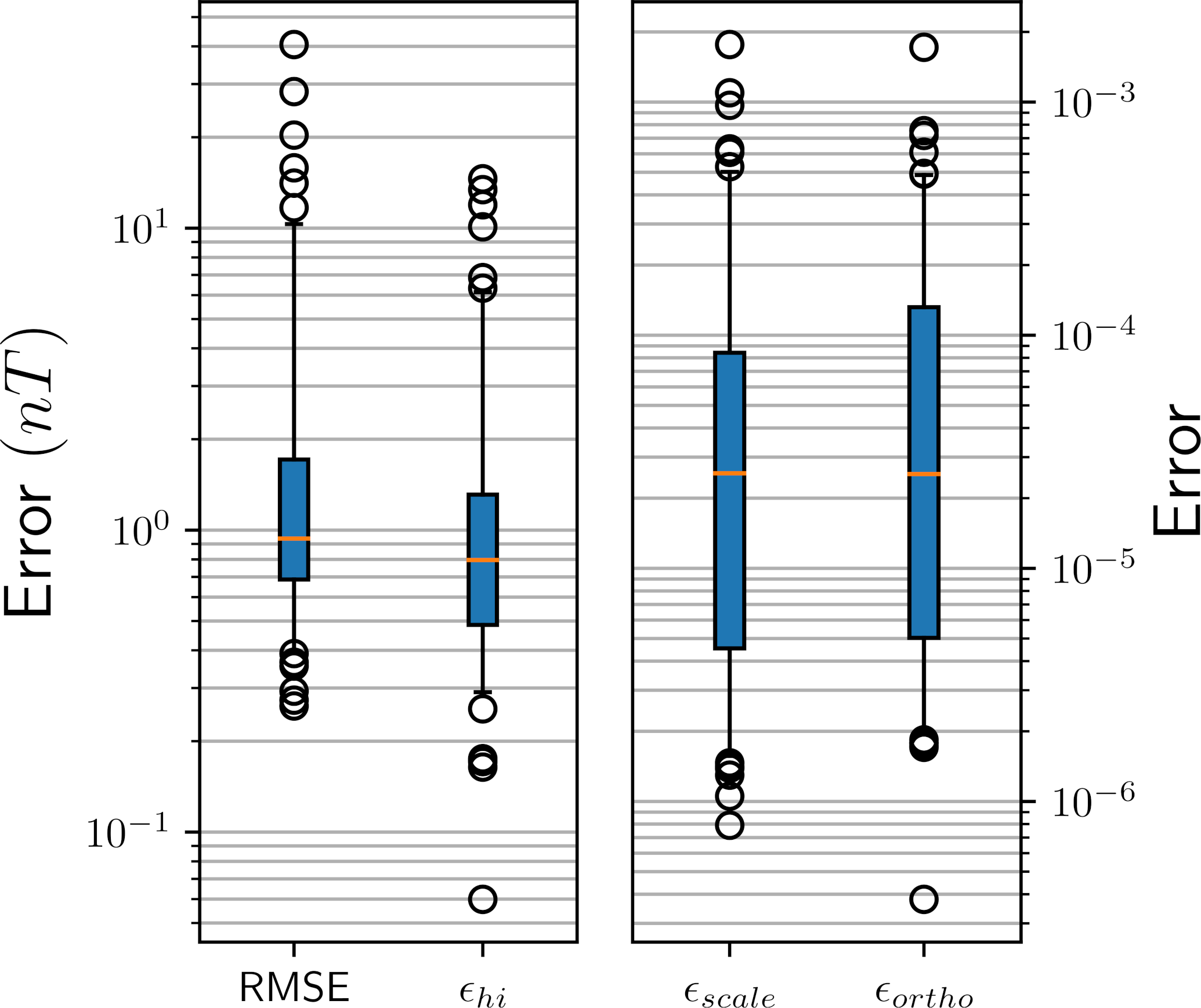}
    \caption{Factor Graph Error Metrics}
    \label{fig:box_plot}{Note: $\epsilon_{scale}$ is unit-less, while $\epsilon_{orth}$ is radians}
\end{figure}

Fig. \ref{fig:box_plot} graphically summarizes the results of the four error metrics in a box plot for the Factor Graph estimation with $\| \boldsymbol{h}_{hi} \| = 5000$ nT.  As seen, the majority of the test runs resulted in sub nanoTesla errors in both the time-varying external field estimate as well as the hard iron contribution.      

The impact of $\|\boldsymbol{h}_{hi}\|$ was assessed by repeating the Monte Carlo simulations with various hard iron sizes ranging from the ideal case of having no contribution up to a magnitude of 5,000 nT which is comparable to magnitudes seen in non-ideal sensor locations in flight test data. The external field was held constant for the simulations comparing performance of the Factor Graph with TWOSTEP and Tolles-Lawson in order to meet the TWOSTEP algorithm assumptions. The results shown in Fig. \ref{fig:h_hi_variation} clearly demonstrate the impact increasing hard iron magnitude has on the estimation accuracy with both TWOSTEP and Tolles-Lawson.  In contrast, hard iron magnitude had minimal effect on the mean hard iron estimate error, although the larger magnitudes did result in slightly larger scatter in the performance.  Despite the increased scatter, the Factor Graph calibration still significantly outperformed TWOSTEP and Tolles-Lawson in simulations with hard iron contributions.


\begin{figure}
    \centering
    \includegraphics{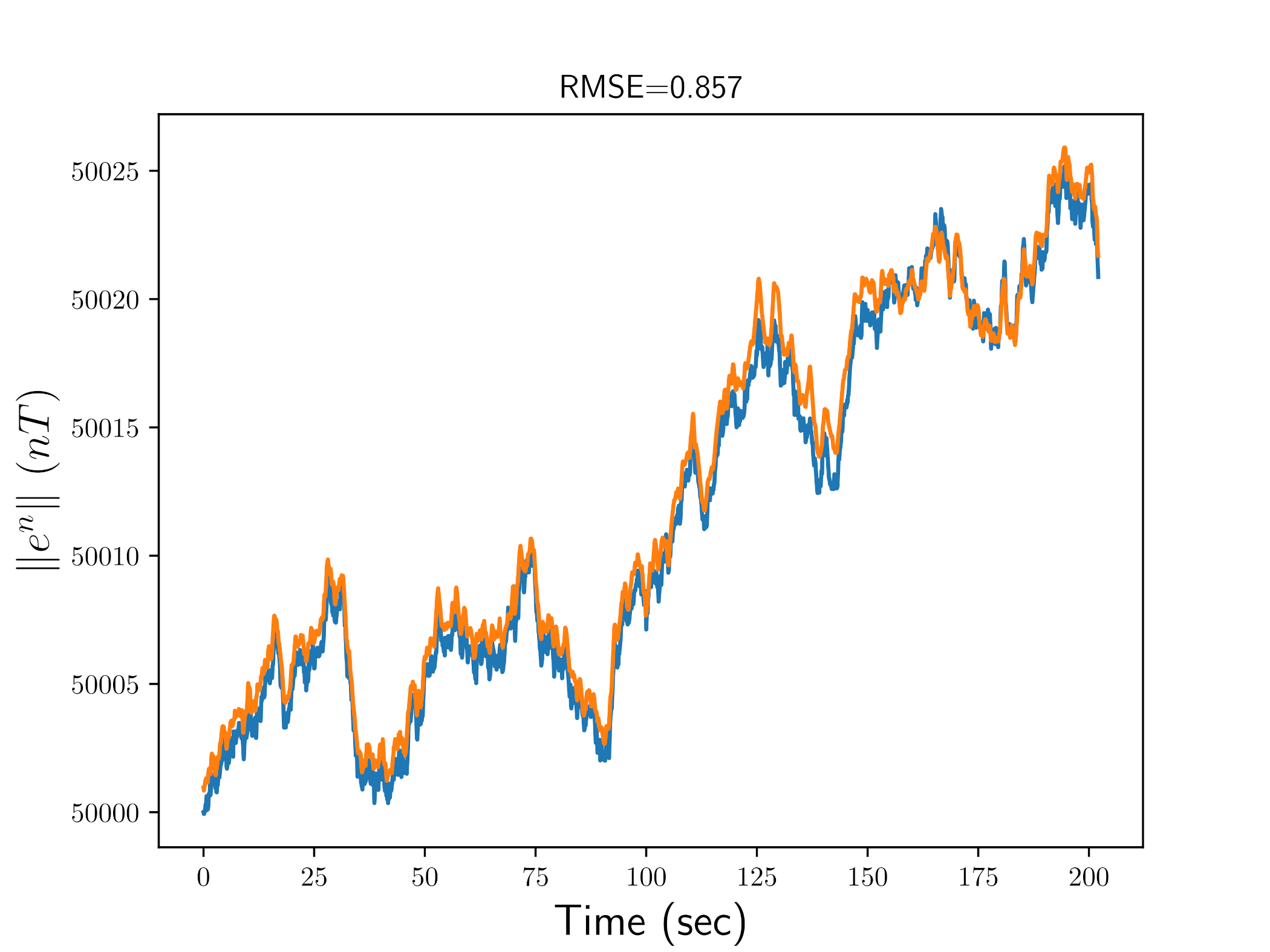}
    \caption{External Field Magnitude Estimate}
    \label{fig:e_est}
\end{figure}

\begin{figure}
    \centering
    \includegraphics{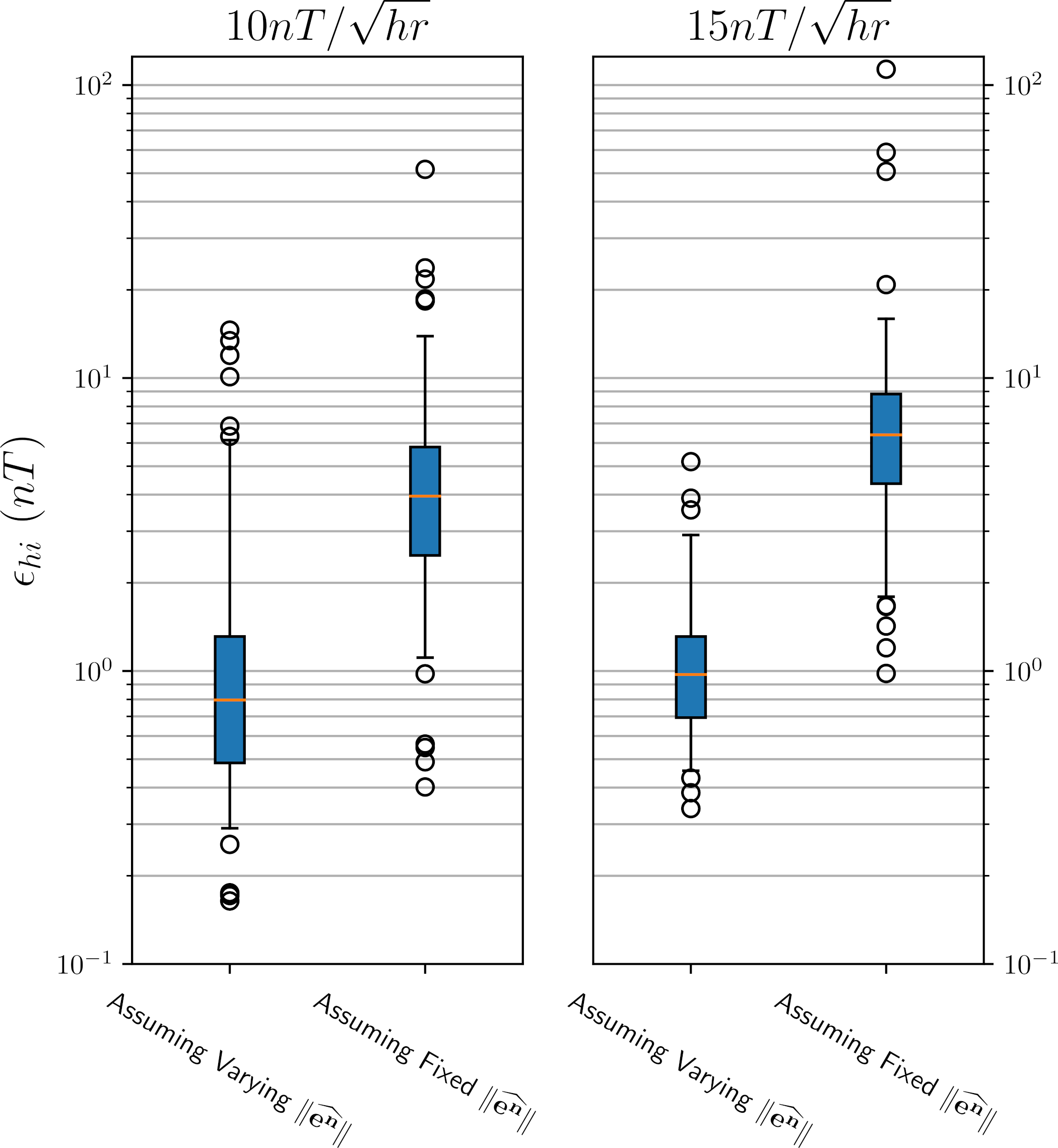}
    \caption{Estimation Performance in a Changing External Field}
    \label{fig:box_plot_comparison}
\end{figure}

In addition to the improved performance of the factor graph when the hard iron effects are large, performance gains were also seen when the external field is not assumed to be constant.  An additional 200 Monte Carlo simulations were conducted where the external field was varied with a random walk noise. The magnitude of the random walk was set to $10 \text{nT}/\sqrt{\text{hr}}$ for half of the runs and $15 \text{nT}/\sqrt{\text{hr}}$ for the remaining half.  A typical external field with random walk is shown in Fig. \ref{fig:e_est} along with the factor graph's estimation of the field.  The ability of the the algorithm to estimate a time-varying external field reduced error in hard iron estimates compared to assuming a constant external field when some variation was actually present.  Fig. \ref{fig:box_plot_comparison} presents two box plots comparing the the results when the external field is assumed to be constant, but isn't, versus assuming the field is not constant.  Fig. \ref{fig:box_plot_comparison} shows the results with the two random walk magnitudes when the algorithm is allowed to estimate the changing field versus assuming a constant external field. The hard iron estimate error is reduced when the constraint of a fixed external field is removed. 

\subsection{Experimental Evaluation} \label{section:Experiement}

The factor graph calibration method was tested in a controlled environment using a scalar and vector magnetometer.  The Geometrics Micro-Fabricated Atomic Magnetometer (MFAM) is a laser pumped cesium  magnetometer and provided scalar measurements during data collections \cite{website:MFAM}. A TwinLeaf magnetoresistive vector magnetometer (VMR) provided vector magnetometer and inertial measurements \cite{website:VMR}.  The attitude of the platform was calculated from the VMR accelerometers, gyros, and vector magnetometer.  Table \ref{tab:sens_param} summarizes the performance characteristics of the utilized sensors.  

\begin{table}[h]
   \renewcommand{\arraystretch}{1.3}
   \caption{Magnetic Sensor Performance}
    \begin{threeparttable}
   \centering
   \begin{tabular}{|p{85pt}|p{65pt}|p{65pt}|}
   \hline
   \bf{Parameter} & \bf{Geometrics MFAM} & \bf{TwinLeaf VMR}  \\
   \hline\hline
   \bf{Sensitivity} $\left(pT/\sqrt{Hz}\right)$   & 5 & 300   \\
   \hline
    \bf{Dead Zone} (degrees)& $\pm 25$\tnote{2} & $N/A$  \\
   \hline
   \bf{Heading Error} (nT) & $\pm 5$\tnote{2} & $N/A$ \\
   \hline
   \bf{Sampling Rate\tnote{1}} (Hz) & 1000 & 200 \\
   \hline
   \end{tabular}
   \begin{tablenotes}
    \small
    \item[1] Sampling rate of sensor; measurements were filtered and down sampled to 10 Hz for implementation 
    \item[1] MFAM sensors oriented for low heading errors
   \end{tablenotes}
   
   \end{threeparttable}
   \label{tab:sens_param}
\end{table}

\subsection{Data Collection}

The VMR and MFAM sensors were mounted on a 3 axis gimbal made of non-magnetic materials such that sensors were mounted near the intersection of the three axes of rotation. This arrangement minimized sensor position changes in the earth frame while performing the calibration rotations.  It was observed that placing the VMR immediately adjacent to the MFAM sensor heads caused interference in the MFAM readings.  Ideally the sensors would have been collocated to minimize any spatial differences in sensed magnetic fields, however it was necessary to maintain approximately four inches between MFAM and VMR sensors to prevent interference in the MFAM readings.

\begin{figure}[htbp]
    \centering
    \begin{subfigure}[b]{.48\textwidth}
        \includegraphics[width=\textwidth, keepaspectratio]{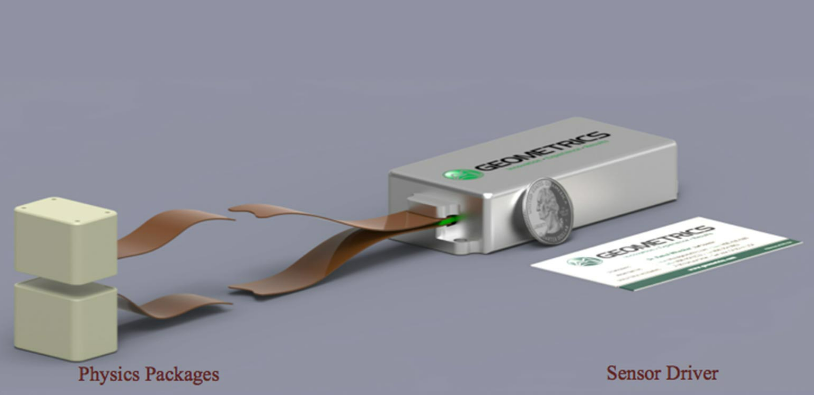}
        \caption{Geometrics MFAM \cite{website:MFAM}}
    \end{subfigure}
    \vfill
    \begin{subfigure}[b]{.48\textwidth}
        \includegraphics[width=\textwidth, keepaspectratio]{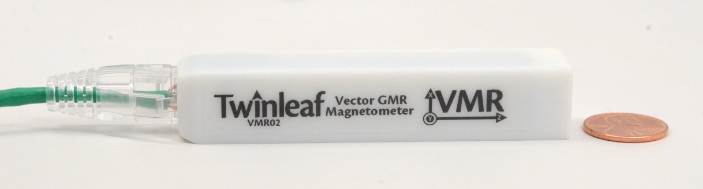}
        \caption{TwinLeaf VMR \cite{website:VMR}}
    \end{subfigure}    
    \caption{Magnetic Sensors}
    \label{fig:my_label}
\end{figure}

The platform was manipulated to move the sensors through a series of pitch, roll, and yaw doublets on four headings approximately 90 degrees apart, as described in Section \ref{sec:sim_setup}.  This profile was chosen to closely mimic maneuvers performed by an aircraft during an airborne calibration flight.

\subsection{Data Collection Results}

The purpose of the real data collections was to determine if the algorithm can effectively calibrate the platform for hard iron contributions.  In order to make this determination, two data sets were collected using the same calibration profile: the first was accomplished with just the test stand and a second after adding permanent magnets to the platform to change the hard iron characteristics.  In order to measure ``truth", data were recorded while the platform was motionless and the permanent magnets were added.  Measuring the change in the VMR magnetic readings before and after the permanent magnets were placed provided an approximate change in the platform's hard iron characteristics.    

Calculating the difference resulted in: 

\begin{equation}
    \Delta\boldsymbol{h}_{hi} \approx \begin{bmatrix}
        -260.97 & -122.07 & -1742.65
    \end{bmatrix} \text{nT}
    \label{eq:delta_hi}
\end{equation}

Calibration data were collected both before and after the addition of the permanent magnets and each were separately used by the calibration routine to estimate the value of the platform's $\boldsymbol{h}_{hi}$. The results of the before and after calibrations were then compared to calculate an estimated hard iron change, $\Delta\boldsymbol{\hat{h}}_{hi}$, from the addition of the magnets. The results when using our factor graph calibration are shown in \eqref{eq:before} - \eqref{eq:epsilon_delta_h_hi}.

\begin{align}
    \label{eq:before} Before: \hat{\boldsymbol{h}}_{hi} &= \begin{bmatrix} -108.17 & -69.46 & 30.99 \end{bmatrix} \text{nT}\\
    After: \hat{\boldsymbol{h}}_{hi} &= \begin{bmatrix} -407.19 & -236.40 & -1651.84 \end{bmatrix} \text{nT}\\
    \Delta\hat{\boldsymbol{h}}_{hi} &= \begin{bmatrix} -299.02 & -166.94 & -1682.83 \end{bmatrix} \text{nT}\\
    \label{eq:epsilon_delta_h_hi}\epsilon &= \begin{bmatrix} -38.05 & -44.87 & 59.82 \end{bmatrix} \text{nT}
\end{align}

These results show the factor graph to be a suitable method for calibrating hard iron effects.  While $\epsilon$ appears large on initial inspection, a significant portion of the results' error could be due to the limitation of the test fixture.  While it was assumed the changes in the VMR measurements were due to the permanent magnet addition, some of the difference is also likely due to a change in platform orientation.  The platform orientation was attempted to be held constant while adding the magnets, however with the size of the external field one tenth of a degree of orientation change would manifest itself as ~100 nT change in the hard iron contribution.     

The same measurement data sets were used with the TWOSTEP and Tolles-Lawson calibration to compare the results. Using the same method to calculate the error from the expected change in hard iron compared to the estimated values resulted in:

\begin{align}
    \epsilon_{TWOSTEP} &= \begin{bmatrix} -15.05 & 27.72 & 546.59 \end{bmatrix} \text{nT}\\
    \epsilon_{Tolles-Lawson} &= \begin{bmatrix} 7.89 & 55.79 & -1390.17 \end{bmatrix} \text{nT} 
\end{align}

The Factor Graph method had the lowest error when using the same measurements to estimate the hard iron characteristics of the platform. 

\begin{align}
    \|\epsilon\| &= 83.9 \text{ nT}\\
    \|\epsilon_{TWOSTEP}\| &= 547.5 \text{ nT}\\
    \|\epsilon_{Tolles-Lawson}\| &= 1391.3 \text{ nT}
\end{align}

\section{Conclusion} \label{section:Conclusion}

This paper demonstrates the suitability of factor graph optimization for the purpose of calibrating platform magnetic fields. The factor graph calibration method presented is 1) significantly more robust to large hard iron effects than prior approaches and 2) can remove the assumption of a fixed external field in which the calibration is being performed. The factor graph method had lower errors when estimating the platform hard iron with both simulated data and when using real data. These attributes makes it well suited for non-ideal sensor installations often necessary in aircraft not intended for magnetic survey work. Although the results were promising, additional work is required to increase the robustness of the proposed method.  The primary shortcoming to address is the the current algorithm was limited to calibrating hard iron effects, whereas other contemporary routines solve for both hard iron and soft iron, and in the case of Tolles-Lawson, eddy currents.  Additionally, we assume the hard iron effects are the same for both the scalar and vector sensors, which is dependent on sensor location.  Our lab experiments mitigated this limitation by adding hard iron effects (permanent magnets) at a location the resulted in similar magnitude changes in both vector and scalar sensors.  Despite this current limitation, the Factor Graph algorithm presents a novel method of platform magnetic calibration suitable for large hard iron fields. Additionally, allowing a time-variant external field allows calibrations in environments that cannot be as tightly controlled as a lab experiment.

\bibliographystyle{ieeetr}
\bibliography{bib}

\end{document}